\documentclass[12pt]{article}
\begin{document}
\setlength{\baselineskip}{0.30in}
\newcommand{\nc}{\newcommand}
\newcommand{\beq}{\begin{equation}}
\newcommand{\eeq}{\end{equation}}
\newcommand{\be}{\begin{eqnarray}}
\newcommand{\ee}{\end{eqnarray}}
\newcommand{\num}{\nu_\mu}
\newcommand{\nue}{\nu_e}
\newcommand{\nut}{\nu_\tau}
\newcommand{\nua}{\nu_a}
\newcommand{\nus}{\nu_s}
\newcommand{\mnus}{m_{\nu_s}}
\newcommand{\taus}{\tau_{\nu_s}}
\newcommand{\nnt}{n_{\nu_\tau}}
\newcommand{\rnt}{\rho_{\nu_\tau}}
\newcommand{\mnt}{m_{\nu_\tau}}
\newcommand{\tnt}{\tau_{\nu_\tau}}
\newcommand{\bi}{\bibitem}
\newcommand{\rar}{\rightarrow}
\newcommand{\lar}{\leftarrow}
\newcommand{\lrar}{\leftrightarrow}
\newcommand{\dm}{\delta m^2}
\newcommand{\sv}{\sin 2\theta}
\newcommand{\so}{\, \mbox{sin}\Omega}
\newcommand{\co}{\, \mbox{cos}\Omega}
\newcommand{\sotil}{\, \mbox{sin}\tilde\Omega}
\newcommand{\cotil}{\, \mbox{cos}\tilde\Omega}
\newcommand{\raa}{\rho_{aa}}
\newcommand{\rss}{\rho_{ss}}
\newcommand{\rsa}{\rho_{sa}}
\newcommand{\ras}{\rho_{as}}
\nc{\app}[3]{{\it  Astropart.\ Phys. }{{\bf #1} {(#2)} {#3}}}
\nc{\ncim}[3]{{\it  Nuov.\ Cim.\ }{{\bf #1} {(#2)} {#3}}}
\nc{\np}[3]  {{\it  Nucl.\ Phys.\ }{{\bf #1} {(#2)} {#3}}}
\nc{\pr}[3]  {{\it  Phys.\ Rev.\ }{{\bf #1} {(#2)} {#3}}}
\nc{\pra}[3] {{\it  Phys.\ Rev.\ A\ }{{\bf #1} {(#2)} {#3}}}
\nc{\prb}[3] {{\it  Phys.\ Rev.\ B\ }{{{\bf #1} {(#2)} {#3}}}}
\nc{\prc}[3] {{\it  Phys.\ Rev.\ C\ }{{\bf #1} {(#2)} {#3}}}
\nc{\prd}[3] {{\it  Phys.\ Rev.\ D\ }{{\bf #1} {(#2)} {#3}}}
\nc{\prl}[3] {{\it  Phys.\ Rev.\ Lett.\ }{{\bf #1} {(#2)} {#3}}}
\nc{\pl}[3]  {{\it  Phys.\ Lett.\ }{{\bf #1} {(#2)} {#3}}}
\makeatletter
\def\alt{\mathrel{\mathpalette\vereq<}}
\def\vereq#1#2{\lower3pt\vbox{\baselineskip1.5pt \lineskip1.5pt
\ialign{$\m@th#1\hfill##\hfil$\crcr#2\crcr\sim\crcr}}}
\def\agt{\mathrel{\mathpalette\vereq>}}

\newcommand{\eq}{{\rm eq}}
\newcommand{\tot}{{\rm tot}}
\newcommand{\M}{{\rm M}}
\newcommand{\coll}{{\rm coll}}
\newcommand{\ann}{{\rm ann}}
\makeatother

\begin{center}
\vglue .06in
{\Large \bf { Neutrino oscillations in the early universe.
Resonant case.
}}
\bigskip
\\{\bf A.D. Dolgov
\footnote{Also: ITEP, Bol. Cheremushkinskaya 25, Moscow 113259, Russia.}
 \\[.05in]
{\it{INFN section of Ferrara\\
Via del Paradiso 12,
44100 Ferrara, Italy}
}}
\end{center}

\begin{abstract}

Lepton asymmetry generated in the early universe by neutrino oscillations 
into sterile partners is calculated. Kinetic equations are analytically
reduced to a simple form that permits an easy numerical treatment. 
Asymptotic values of the asymmetry are at the level of 0.2-0.3 and are
reasonably close to those obtained by other groups, though the approach to 
asymptotics in some cases is noticeably slower. No chaoticity is observed.

\end{abstract}   

\section{Introduction.} \label{sec:intr}

Neutrino oscillations in the early universe might possess a very
interesting 
property, if active neutrinos ($\nu_a=\nue,\num,\nut$)
are mixed with sterile ones ($\nus$). Refraction index
of neutrinos in the cosmic plasma depends upon the cosmological charge 
asymmetry of the plasma, $\eta$, which is normally quite small, 
$\eta = 10^{-9}-10^{-10}$. Because of this dependence the transformation
of $\nua$ into $\nus$ might be slightly more favorable than the 
transformation of the corresponding antineutrinos (or vice versa depending 
upon the sign 
of the initial asymmetry). The feedback effect is positive and
leads to a further increase of asymmetry making, say, $\nua \rar \nus$
transformation more and more efficient in comparison with
$\bar\nua \rar \bar\nus$. 
Of course the total leptonic charge of active plus sterile neutrinos is
conserved but refraction index depends only on charge asymmetry in the 
sector of active neutrinos, and the lepton asymmetry in active
sector could strongly increase.

The effect of asymmetry generation takes place only for a 
sufficiently small mixing; i.e. for $\dm \sim 1$ eV$^2$ it is 
roughly bounded by
$(\sv)^2 \leq\, (\sim10^{-3})$, where $\theta$ is the vacuum mixing
angle. For large mixings active and sterile states of both neutrinos 
and antineutrinos would quickly reach thermally equilibrium values, they 
would become equally populated, and this prevents from generation of 
a large asymmetry. For a positive mass difference between 
$\nus$ and $\nua$ ($\dm >0$) the asymmetry would remain small 
also in the case of small mixing. However for $\dm <0$ the resonance 
MSW-transition~\cite{mikheev85,wolfenstein78} might take place in the 
primeval plasma and this effect could compensate a smallness of vacuum
mixing and produce a considerable lepton asymmetry in the sector of
active neutrinos.

The instability with respect to generation of lepton asymmetry by neutrino
oscillations was noticed in ref.~\cite{barbieri91} but it was concluded
there, on the basis of simplified arguments, that the rise of asymmetry
was terminated when it was still quite small. 
This conclusion was reconsidered in ref.~\cite{foot96} (see also
refs.~\cite{foot97a}-\cite{foot99}) where it was argued that a very large
asymmetry, close to 1, may be generated by the oscillations. This
result was questioned in ref.~\cite{dolgov00} where it 
was claimed that back reaction effects are very strong, they
significantly slowed down the rise of asymmetry, and the latter
could reach only the value around $10^{-6}-10^{-4}$ for reasonably
small mass differences. However certain drawbacks of the approach 
of the paper~\cite{dolgov00} were indicated in 
ref.~\cite{dibari00b} (see also~\cite{buras00a,buras00b})
and though not all of them were relevant, it would be 
enough to have one weak point to destroy a conclusion. 

Since the result of a large asymmetry generation is very interesting
and important (in particular, for the big bang nucleosynthesis, BBN) it 
is worthwhile to make an independent calculation of the effect. In this
work kinetic equations governing neutrino oscillations in the early
universe are analytically transformed to a rather simple form which
after well controlled approximate simplifications permits an easy 
numerical solution. 

The results of this work are reasonably close to those found in 
refs.~\cite{foot97a,foot97b,foot99,dibari00b} for the case 
of $(\nu_{\mu,\tau}-\nus)$-oscillations and approximately 2-3 times
smaller than the asymmetry found in ref.~\cite{dibari00c} for
$(\nue-\nus)$-oscillations in the temperature range essential for
BBN. 

Another interesting issue related to the asymmetry generation is 
its possible chaoticity so that the sign of the final large value 
of the lepton asymmetry is a rapidly oscillating function of the 
oscillation parameters or of a small variation of the initial value of 
the cosmological charge asymmetry~\cite{shi96}-\cite{enqvist00}. 
According to the calculations of the present paper, no chaoticity 
exists at least in the parameter range considered below, for which a
large lepton asymmetry is generated. 

\section{Kinetic equations.} \label{sec:kineq}

We assume that a non-negligible mixing exists only between one active and
one sterile neutrino, so that the neutrino state is described by 
$2\times 2$\,-\,density matrix $\rho$. Its evolution is governed by 
the usual equation:
\be 
i\dot \rho = \left[ {\cal H}, \rho \right]
\label{dotrho}
\ee
where ${\cal H}$ is the neutrino Hamiltonian. It contains the free part:
which is diagonal in the mass basis:
\be
{\cal H}_{free} = {\rm diag} \left[ \sqrt{p^2 + m_a^2},\,\,\sqrt{p^2 + 
m_s^2} \,\right]
\label{hfree}
\ee
and the part that describes interaction with medium which looks simpler
in the flavor basis. The interaction Hamiltonian contains first order 
terms given by refraction index (or effective potential of neutrinos). 
It was originally calculated in ref.~\cite{notzold88} and consists
of two terms (see also discussion in papers~\cite{barbieri90,barbieri91}:
\be
V_{eff}^a =
\pm C_1 \eta^{(a)} G_FT^3 + C_2^a \frac{G^2_F T^4 E}{\alpha} ~,
\label{nref}
\ee
where $E$ is the neutrino energy, $T$ is the temperature of the
plasma, $G_F=1.166\cdot 10^{-5}$ GeV$^{-2}$ is the Fermi coupling
constant, $\alpha=1/137$ is the fine structure constant, and the signs
``$\pm$'' refer to anti-neutrinos and neutrinos respectively. 
According to ref.~\cite{notzold88} the
coefficients $C_j$ are: $C_1 \approx 0.95$, $C_2^e \approx 0.61$ and
$C_2^{\mu,\tau} \approx 0.17$.  These values are true in the case of
thermal equilibrium, otherwise these coefficients are some
integrals over the distribution functions. 
The contributions to $\eta^{(a)}$ from different
particle species are the following:
\be
\eta^{(e)} = 
2\eta_{\nue} +\eta_{\num} + \eta_{\nut} +\eta_{e}-\eta_{n}/2 \,\,\,
 ( {\rm for} \,\, \nue)~,
\label{etanue} \\
\eta^{(\mu)} = 
2\eta_{\num} +\eta_{\nue} + \eta_{\nut} - \eta_{n}/2\,\,\,
({\rm for} \,\, \num)~,
\label{etanumu}
\ee
and $\eta^{(\tau)}$ for $\nut$ is obtained from eq.~(\ref{etanumu}) by 
the interchange $\mu \lrar \tau$. The individual charge asymmetries, 
$\eta_X$, are defined as the ratio of the difference between 
particle-antiparticle number densities to the number density of photons:
\be
\eta_X = \left(N_X -N_{\bar X}\right) /N_\gamma
\label{etax}
\ee

The interaction Hamiltonian contains also the so called second order 
terms that describe the loss of coherence due to elastic or inelastic
neutrino scattering and annihilation as well as neutrino production by
reactions in the primeval plasma. For the description of the loss of
coherence it is enough to take into account only imaginary part of
the second order Hamiltonian. Discussion and calculations can be found 
in the papers~\cite{harris78}-\cite{volkas00}. The exact form of the
second
order terms has quite complicated matrix structure, it is non-linear in 
$\rho$, and contains multi-dimensional integrals over phase space (which
can be reduced down to two dimensions). Their explicit expressions can be 
found e.g. in refs.~\cite{dolgov81,sigl93}. However it is very difficult,
to solve kinetic equations for density matrix with the exact
expressions for the second order terms, especially in the resonance 
case. (An approximate solution in non-resonance case is found in 
ref.~\cite{dolgov00b}). Hence one usually mimic the exact expression by
the linear ``poor man'' substitution:
\be 
i\dot\rho = ...(i/2) \left\{ \Gamma,\, \rho - \rho_{eq} \right\}...
\label{dotrhogam}
\ee
where curly brackets mean anti-commutator and the matrix $\Gamma$ 
effectively describes neutrino scattering and annihilation. It has the
only non-vanishing entry at $aa$-corner in the flavor basis, which is 
taken as
\be
\Gamma_{aa}\equiv \Gamma_0^a = C^a_\Gamma G^2_F T^4 E
\label{gammaaa}
\ee
where $ C^a_\Gamma$ is a constant. According to
reference~\cite{enqvist92b} 
$C^e_\Gamma =1.27$ and $C^{(\mu,\tau)}_\Gamma=0.92 $, while 
according to~\cite{dolgov00}, where slightly more accurate calculations 
were done, $C^e_\Gamma =1.13 $ and $C_\Gamma^{(\mu,\tau)}=0.79$. 
This difference and even the absolute value of $\Gamma_{aa}$
are not essential for the magnitude of rising asymmetry because
$\Gamma$ remains in some sense small and can be neglected (see below).
Even the use of the approximate expression for the coherence breaking
terms~(\ref{dotrhogam}) in contrast to the exact collision integral
which changes the results in the case of non-resonance oscillations by
an order of magnitude~\cite{dolgov00b}, have a very weak impact on the 
value of the generated lepton asymmetry in the resonance case.

The matrix $\rho_{eq}$ is equal to $f_{eq}(\mu)\, I$ where $I$ is the
unit matrix and and $f_{eq}(E/T,\mu)$ is the equilibrium Fermi
distribution
function:
\be
f_{eq}={1\over  \exp\left[ \left(E-\mu\right)/T\right] +1}~.
\label{feq}
\ee
where $\mu$ is neutrino chemical potential. In what follows we will take
$\mu =0$, though in 
refs.~\cite{foot96}-\cite{foot99},\cite{dibari00b} it was
taken equal to the running value determined by the generated
lepton asymmetry. 
However the final result for the asymmetry is not sensitive to this
assumption (see discussion in section~\ref{sec:discussion}.

In the Friedman-Robertson-Walker universe the matrix elements of
$\rho$ satisfy the following equations:
\be
i(\partial_t -Hp\partial_p) \rho_{aa}
&=& F_0(\rho_{sa}-\rho_{as})/2 -i \Gamma_0 (\rho_{aa}-f_{eq})~,
\label{dotrhoaa} \\
i(\partial_t -Hp\partial_p)  \rho_{ss}
&=& -F_0(\rho_{sa}-\rho_{as})/2~,
\label{dotrhoss} \\
i(\partial_t -Hp\partial_p) \rho_{as} &=&
W_0\rho_{as} +F_0(\rho_{ss}-\rho_{aa})/2-
i\Gamma_0 \rho_{as}/2 ~,
\label{dotrhoas}\\
i(\partial_t -Hp\partial_p) \rho_{sa} &=& -W_0\rho_{sa} -
F_0(\rho_{ss}-\rho_{aa})/2- i\Gamma_0/2 \rho_{sa}  ~,
\label{dotrhosa}
\ee
where $a$ and $s$ mean ``active'' and ``sterile'' respectively, 
\be
F_0=\dm \sin 2\theta / 2E,
\label{f0} \\
W_0= \dm\cos 2\theta /2E + V_{eff}^a,
\label{w0}
\ee
$H=\sqrt{8\pi \rho_{tot}/3M_{Pl}^2}$ is the Hubble parameter, $M_{Pl}$
is the Planck mass, and $p$ is the neutrino momentum. The neutrinos are
assumed to be very light, so that $p=E$. The total cosmological energy
density is taken as $\rho_{tot} = 10.75\pi^2 T^4/30$. 
This corresponds to photons, $e^\pm$-pairs, and
three types of neutrinos in thermal equilibrium with equal temperatures 
$T$. This approximation is quite good when $T$ is larger than a few 
hundred keV. At smaller temperatures $e^+e^-$-annihilation heats up
electromagnetic component of the plasma and the temperatures of photons
and neutrinos become different.  

We have prescribed sub-index ``0'' to the
coefficient functions in eqs.~(\ref{dotrhoaa}-\ref{dotrhosa}) to 
distinguish them from the same ones divided by $Hx$ 
(see eqs.~(\ref{s'}-\ref{coef}) below).

The anti-neutrino density matrix, $\bar \rho$,
satisfies the similar set of equations
with the opposite sign of the antisymmetric term in $V_{eff}^a$ and with
a slight difference in damping factors $\bar\Gamma_0$ which is 
proportional to the lepton asymmetry. 

It is convenient to introduce new variables:
\be
x=m_0 R(t)\,\, {\rm and}\,\,   y=p R(t)~,
\label{xy}
\ee
where $R(t)$ is the cosmological scale factor so that $H=\dot
R/R$ and $m_0$ is an arbitrary mass (just normalization), taken as
$m_0 =1$ MeV. One may approximately assume that
$\dot T = -HT$, and correspondingly $R=1/T$. In terms of these
variables the differential operator $(\partial_t -Hp\partial_p)$
transforms to $Hx\partial_x$. We will normalize the density matrix
elements to the equilibrium distribution with zero chemical potential:
\be
\rho_{aa} &=& f_{eq}(y)\, [1+a(x,y)],\,\, 
\rho_{ss} = f_{eq}(y)\, [1+s(x,y)]~, \\
\label{rhoaa}
\rho_{as} &=& \rho_{sa}^* = f_{eq}(y)\,[h(x,y)+i l(x,y)]~,
\label{hil}
\ee
Other authors find it convenient to express the density matrix
in terms of Pauli matrices and the polarization vector, $\vec P =
(P_x, P_y, P_z)$, in such a way that:
\be
\rho \equiv \frac{P_0}{2} \left[ 1 + \vec P \cdot \vec \sigma \right]~,
\label{rhosigma}
\ee
The relation between different functions are $P_0P_x = 2f_{eq}h$, 
$P_0P_y = -2f _{eq} l$,
$P_0P_z = f_{eq} (a-s)$ and $P_0 = f_{eq} (2 + a+s)$. In particular,
$P_z = 1$ means that all the neutrinos are active, $\nu_e,\, \num$,
or $\nut$. 

Let us now divide both sides of equations~(\ref{dotrhoaa}-\ref{dotrhosa}) 
by $Hx$ and denote the corresponding coefficient 
functions~(\ref{f0}-\ref{w0}) with sub-index ``1'', e.g. 
$W_1 = W_0 /Hx = U_1 \pm V_1 Z$, etc and thus we get: 
\be
U_1 &=& 1.12\cdot 10^9\cos 2\theta \delta m^2 {x^2\over y}+
\tilde C_2^a\, {y\over x^4},
\label{u1}\\
V_1 &=&\frac{30}{x^2},
\label{v1}\\
Z &=& 10^{10}\left[ {\eta_{o} \over 12}
+ \int_0^\infty \frac{dy}{8 \pi^2} ~y^2 f_{eq}(y) ~(a-\bar a)\right]~,
\label{Z}
\ee 
where $\dm$ is expressed in eV$^2$ (here and below),
$\tilde C_2^e \approx 26$ and $\tilde C_2^{\mu,\tau}\approx 7$ and 
$\eta_o$ is the charge asymmetry of all particles except for $\nua$ 
defined in accordance with eqs.~(\ref{etanue},\ref{etanumu}). 
The normalization 
of the charge asymmetry term~(\ref{Z}) is rather unusual and to 
understand the numerical values of the coefficients one should keep
in mind the following. The coefficient $C_2$ in eq.~(\ref{nref}) is
found for the standard normalization of charge asymmetry with respect 
to the present day photon number density which differs from that in the
early universe by the well known factor 11/4, related to the increase of
photon number by $e^+e^-$-annihilation. On the other hand, lepton 
asymmetry, $L_{\nua}$, induced by neutrino oscillations, which is 
calculated in most of papers is normalized to the number density of 
photons that are in thermal equilibrium with neutrinos, so the factor 
11/4 is absent. The photon number density is equal to 
$N_\gamma = 2\zeta (3) T^3_\gamma /\pi^2$ ($\zeta(3)\approx 1.2$). 
The charge asymmetry of neutrinos is
\be
\eta_\nu = {1\over 4\zeta (3)} \left({T_\nu \over T_\gamma}\right)^3
\int dy y^2 (\rho_{aa} - \bar\rho_{aa})
\label{etanu}
\ee
so that $\eta_{\nua} = 4 L_{\nua}/11$. The quantity $Z$ introduced in 
eq.~(\ref{Z}) differs from $L_{\nua}$ by the factor 
$2\cdot 10^{-10}\pi^2 /\zeta(3)$:
\be
L = 16.45\cdot 10^{-10}\, Z.
\label{lz}
\ee
The factor $10^{-10}$ is chosen so that initially $Z = O(1)$. 
Noting that the charge asymmetry of neutrinos under study enters 
expressions (\ref{etanue},\ref{etanumu}) with coefficient 2 one obtains
the coefficient $1/11.96\approx 1/12$ in eq.~(\ref{Z}).

Now, following ref.~\cite{dolgov00}, we will introduce one more new 
variable $q=\xi_a x^3 $ in such a way that $U_1$, eq.~(\ref{u1}),
vanishes at $q=y$ independently of the values of the oscillation
parameters: 
\be
U_1 = 1.12\cdot 10^9\,\cos 2\theta\, |\delta m^2|\, {x^2 \over y} \,
\left[ -1 + \left({y\over q}\right)^2 \right]
\label{uq}
\ee
(it is assumed that $\dm<0$). The coefficients $\xi_a$ are
\be
\xi_e = 6.63\cdot 10^3 \left(|\dm| \cos 2\theta\right)^{1/2}, \,\,\,
\xi_{\mu,\tau} = 1.257\cdot 10^4 \left(|\dm| \cos 2\theta\right)^{1/2}
\label{xiemu}
\ee
In what follows we assume that $\sv \ll 1$ and $\cos 2\theta\approx 1$.

Written in terms of the variable $q$ the system of basic kinetic 
equations takes a very simple form~\cite{dolgov00}:
\be
s' &=& -(K_a/y) \sv\, l
\label{s'} \\
a' &=& (K_a/y) \left( \sv\, l - 2\gamma\, a \right)
\label{a'}\\
h' &=& (K_a/y) \left( W l - \gamma \, h\, \right)
\label{h'} \\
l' &=& (K_a/y) \left[\sv\,(s-a)/2 -Wh - \gamma\, l \right]
\label{l'}
\ee
where $K_a = 1.12\cdot 10^9\,\cos 2\theta \, |\dm| /3\xi_a$, so
$K_e = 5.63\cdot 10^4 (|\dm|\,\cos 2\theta)^{1/2}$ and
$K_{\mu,\tau} = 2.97\cdot 10^4 (|\dm|\,\cos 2\theta)^{1/2}$.
In what follows we use the limit $K_a \gg 1$. This permits to make
accurate analytic calculations. A large magnitude of this coefficient
reflects a large frequency of neutrino oscillations with respect to
other essential time scales. Its large value makes numerical solution
very difficult but allows to make accurate analytical estimates.

The coefficient functions in these equations have the form:
\be
W = U \pm y\,V\,Z, \,\,\, U = y^2\,q^{-2} -1, \nonumber \\
V = b_a\,  q^{-4/3}, \,\,\, \gamma = \epsilon_a\, y^2 \,q^{-2}
\label{coef}
\ee
where the signs $''-''$ or  $''+''$ in $W$ refer to neutrinos and
antineutrinos respectively; 
$b_e = 3.3\cdot 10^{-3}(|\dm|\,\cos 2\theta)^{-1/3}$,
$b_{\mu,\tau} = 7.8\cdot 10^{-3}(|\dm|\,\cos 2\theta)^{-1/3}$, 
and $\epsilon_a$ are small coefficients, 
$\epsilon_e \approx 7.4\cdot 10^{-3}$
and $\epsilon_{\mu,\tau} \approx 5.2\cdot 10^{-3}$. Their exact 
numerical value is not important.
It is noteworthy that the charge asymmetric term in $W$ comes with a
very large coefficient if expressed in terms of $L$, 
$VZ \sim 10^7 q^{-4/3} L$, while in all other possible places 
$L$ (or chemical potential) enters with the coefficient of order 1.

It follows from eqs.~(\ref{s'}-\ref{l'}) that
\be
\partial_q \left( a^2 + s^2 + 2h^2 +2l^2 \right) =
-4\gamma (K/y)\,\left( a^2 +h^2 +l^2\right)
\label{daslh}
\ee
so that the quantity in the r.h.s. may only decrease.

\section{ Solution of kinetic equations.} \label{sec:sol}

One can solve analytically the last two kinetic equations 
(\ref{h'},\ref{l'}) with respect to $h$ and $l$ in terms of $a$ and $s$:
\be
l (q,y)=- ({K \sv / 2y}) \int_{q_{in}}^q dq_1 
\left[ a(q_1) - s(q_1)\right] e^{-\Delta \Gamma} \cos \Delta \Phi,
\label{lqy} \\
h (q,y)=- ({K \sv / 2y}) \int_{q_{in}}^q dq_1 \left[ a(q_1) -
s(q_1)\right]
e^{-\Delta \Gamma} \sin \Delta \Phi,
\label{hqy}
\ee
where $q_{in}$ is the initial ``moment'' $q$ from which the system started
to evolve, $\Delta \Gamma = \Gamma (q,y) - \Gamma (q_1, y)$,
$\Delta \Phi = \Phi (q,y) - \Phi (q_1, y)$, and
\be
\partial_q \Gamma= K\gamma /y,\,\,\, 
\partial_q \Phi= KW /y.
\label{dgammaw}
\ee
We rewrite the first two equations (\ref{s'},\ref{a'}) in terms of 
$\sigma = a + s $ and $\delta = a-s$:
\be
\sigma' &=& -(K\,\gamma /y) \, (\sigma + \delta)
\label{sigma'} \\
\delta' &=&  (2K\sv /y)\,l -(K\,\gamma /y) \, (\sigma + \delta)
\label{delta'}
\ee
The first of these equations can be solved for $\sigma$:
\be
\sigma (q,y) = \sigma_{in}(y)\, e^{-\Gamma(q,y) +\Gamma_{in}(y)} -
{K\over y} \int^q_{q_{in}} dq_1  e^{-\Delta \Gamma} \gamma (q_1,y)
\delta (q_1,y)  
\label{sigma}
\ee
The first term in this expression, proportional, to the initial value
$\sigma_{in}$ is exponentially quickly ``forgotten'' and we obtain
the following equation that contains only $\delta$ (and another unknown
function, integrated charge asymmetry $Z(q)$ that is hidden in the 
phase factor $\Delta \Phi$):
\be
\delta' (q,y) &=&  -{K\,\gamma(q,y) \over y} \, \delta +
{K^2\,\gamma (q,y) \over y^2} \int^q_{q_{in}} dq_1 
e^{-\Delta \Gamma} \gamma (q_1,y) \delta (q_1,y)\\ 
&-&\left({K \sv \over y}\right)^2  \int_{q_{in}}^q dq_1 
\delta (q_1,y) e^{-\Delta \Gamma} \cos \Delta \Phi
\label{deltaf'}
\ee
Up to this point this is an exact equation (with the omitted initial 
value of $\sigma$ which contribution is exponentially small). There is 
also an uncertainty related to the choice of the form of $\rho_{eq}$ in 
eq.~(\ref{dotrhogam}) either with zero or non-zero chemical potential,
see eq.~(\ref{feq}). We have chosen here $\mu =0$ and, as is argued in
sec.~\ref{sec:discussion}, the choice $\mu \neq 0$ does not lead to a 
noticeably different results. This ambiguity could be rigorously
resolved if one uses exact collision integrals instead of 
eq.~(\ref{dotrhogam}). This will be discussed elsewhere. 

Let us take now similar equation for antineutrinos and consider the 
sum and difference of these two equations for charge symmetric and
antisymmetric combinations of the elements of density matrix,
$\Sigma = \delta +\bar \delta$ and $\Delta =\delta -\bar \delta$.
The equations have the following form: 
\be
\Delta' + {K \gamma \over y} \Delta &=& 
{K^2 \gamma \over y^2} \int_{q_{in}}^q dq_1\, e^{-\Delta \Gamma}
\gamma_1 \, \Delta_1 \nonumber \\
&-&\left({K \sv \over y}\right)^2 \int_{q_{in}}^q dq_1\,e^{-\Delta \Gamma}
\left( \Sigma_1 {c-\bar c\over 2} +\Delta_1 {c+\bar c\over 2} 
\right) 
\label{Delta'}\\
\Sigma' + {K\gamma \over y} \Sigma &=& 
{K^2 \gamma \over y^2} \int_{q_{in}}^q dq_1\, e^{-\Delta \Gamma}
\gamma_1 \, \Sigma_1 \nonumber \\
&-&\left({K \sv \over y}\right)^2 \int_{q_{in}}^q dq_1\,e^{-\Delta \Gamma}
\left( \Sigma_1 {c+\bar c\over 2} +\Delta_1 {c-\bar c\over 2} 
\right) 
\label{Sigma'}
\ee
where sub-1 means that the function is taken at $q_1$, e.g. 
$\gamma_1 = \gamma (q_1,y)$, etc; $c = \cos \Delta \Phi$, and
$\bar c = \cos \Delta \bar\Phi$.
Using expressions~(\ref{coef}, \ref{dgammaw}) we find:
\be
{c-\bar c \over 2} = \sin \left[ K \left(q-q_1\right)
\left(-{1 \over y} + {y \over q \,q_1}\right)\right]\,
\sin \left[K\, \int_{q_1}^q dq_2 V(q_2)\,Z(q_2) \right]
\label{cminusc}\\
{c+\bar c \over 2} =\cos \left[ K \left(q-q_1\right)
\left(-{1 \over y} + {y \over q \,q_1}\right)\right]\,
\cos \left[K\, \int_{q_1}^q dq_2 V(q_2)\,Z(q_2) \right]
\label{cplusc}
\ee 
Here $V(q)$ is given by expression~(\ref{coef}) and does not depend on
$y$.

At this stage we will do some approximations to solve the 
system~(\ref{Delta'},\ref{Sigma'}). First, let us consider the terms 
proportional to $\gamma$. They are definitely not important at large 
$q$ (or low temperature). Let us estimate how essential are they at
low $q$ ($q \sim 1$). Integrating by parts the first term in the r.h.s.
of eq.~(\ref{Delta'}), using expression~(\ref{dgammaw}) and
neglecting exponentially small contribution of the initial value, 
we find:
\be
{K\gamma \over y} \Delta - 
{K^2 \gamma \over y^2} \int_{q_{in}}^q dq_1\, e^{-\Delta \Gamma}
\gamma_1 \, \Delta_1 ={K \gamma \over y} 
\int^q_{q_{in}} dq_1 e^{-\Delta\Gamma}\, {d\Delta_1 \over dq_1}
\label{byparts}
\ee
The remaining integral can be easily evaluated in the limit of
large $K\gamma /y = K\epsilon y/q^2$. Indeed, 
$\Delta \Gamma = K \epsilon y (q-q_1)/q\,q_1$ and for $q\leq 1$ the
coefficient in front of the exponential $(q-q_1)$ is larger than 400 for
$\nue$ and 300 for $\num$ and $\nut$. So the integral strongly sits on
the upper limit and, together with the coefficient in front, it gives
just $\delta' (q)$. Thus, when the $\gamma$-terms are large they simply
double the coefficient of $\Delta'$ in eq.~(\ref{Delta'}): 
$\Delta' \rar 2\Delta'$. A possible loophole in this argument is a very
strong variation of the integrand, much stronger than that given by
$\exp (\Delta \Gamma)$. However one can check from the solution found
below that this is not the case.

Thus the role of $\gamma$ terms in eq.~(\ref{Delta'}) is rather mild,
they could only change the coefficient in front of $\Delta'$ from 1 to 
2, and become negligible for large $q$ where the bulk of asymmetry is
generated. So let us neglect these terms in the equation. This 
simplification does not have a strong impact on the solution. 

Let us make one more approximate assumption, namely let us neglect the
second term, proportional to $\Delta_1$, in the last integral of the
r.h.s. of eq.~(\ref{Delta'}). Initially $\Sigma =2$ and 
$ \Delta = 10^{-9}-10^{-10}$ and the neglect of $\Delta$ in comparison
with $\Sigma$ is a good approximation, at least at initial stage. We 
will check the validity of this assumption after we find the solution. 
And last, we assume that $\Sigma$ changes very slowly 
$\Sigma \approx \Sigma_{in} = 2$. Justification for the latter is a
smallness of the mixing angle, $\sv \sim 10^{-4}$. In the limit of
zero mixing, the solution of eq.~(\ref{Sigma'}) is $\Sigma = const$.
We will relax both these assumptions in sec.~\ref{sec:br}.

As a last step we need to find a relation between 
$\Delta= a-s-\bar a+\bar s$ and charge
asymmetry $Z$. To this end one may use the conservation of the total
leptonic charge:
\be
\int^\infty_0 dy\, y^2\, f_{eq}(y) \left( a+s-\bar a -\bar s\right)
= const
\label{cl}
\ee
Using this conservation law we find:
\be
10^{10}\,{d\over dq} \left[\int_0^\infty dy\,y^2 \,f_{eq}(y) 
\Delta (q,y) \right] = 16\pi^2 {dZ\over dq}
\label{deltaz}
\ee

Keeping all these assumptions in mind we can integrate both sides of
eq.~(\ref{Delta'}) with $dy y^2f_{eq}(y)$ and obtain a closed 
ordinary differential equation for the asymmetry $Z(q)$, valid in the
limit of large $K$. Integration over $y$ gives:
\be
Z'(q) &=& {10^{10} K^2 (\sv)^2 \over 8\pi^2} 
\int_0^\infty dy f_{eq}(y) \nonumber \\
&&\int_{q_{in}}^q dq_1\exp \left[ - {\epsilon y \zeta \over qq_1} \right]
\sin\left[ \zeta \left({1\over y}-{y\over qq_1}\right)\right]
\sin\left[bK\int_{q_1}^q dq_2 {Z(q_2) \over q_2^{4/3}}\right]
\label{Z'}
\ee
where $b$ is defined in eq.~(\ref{coef}) and $\zeta = K(q-q_1)$. 
Integration over $y$ here can be done explicitly and the result is
expressed through a real part of a sum of certain Bessel functions of 
complex arguments. To do that one has to expand 
\be
f_{eq} = \sum_n (-1)^{n+1} \exp (-ny)
\label{feqexp}
\ee
and integrate each term analytically~\cite{gradshtein94}.
It can be seen from the result (it is more or 
less evident anyhow) that the integral over $q_1$ is saturated in the 
region $\zeta \sim 1$. So that we can take $qq_1 \approx q^2$ and
\be
K\int_{q_1}^q dq_2 {Z(q_2) q_2^{-4/3}} \approx \zeta Z(q)/q^{4/3}
\label{kint}
\ee
The correction in this expression
is of the order of $Z' (q) /K$. It can be checked, using the solution 
obtained below, that the correction terms are indeed small.

Keeping this in mind we can take the integral over $\zeta$ in
the r.h.s. of eq.~(\ref{Z'}) . To ensure convergence we proceed as 
follows. First, we expand $f_{eq} (y)$ in accordance with 
eq.~(\ref{feqexp}). Each term of the series is non-singular in   
the complex $y$-plane. After that we can
rotate the contour in the complex $y$-plane
to imaginary axis, clockwise for the $\exp [i\zeta (1/y -y/q^2)]$-part
of $\sin  [\zeta (1/y -y/q^2)]$, ($y\rar - iy$), and counter-clockwise 
for the complex conjugate part ($y\rar iy$). Both terms give
$\exp [-\zeta (1/y +y/q^2)]$, so the integral over $\zeta$ is 
exponentially converging for each term in the series~(\ref{feqexp})
and the resulting series is convergent as well:
\be
Z'(q) &=& {10^{10}K(\sv)^2 \over 8\pi^2} \sum_n (-1)^{n+1}
\int_0^\infty {dy\over 4i} \nonumber \\
&&\left[ e^{iny} \left( {1\over {1\over y} + {y(1-i\epsilon)\over q^2} -
i\phi} - {1\over {1\over y} + {y(1-i\epsilon)\over q^2} +
i\phi} \right)  \right. \nonumber \\
&& \left.
+e^{-iny} \left( {1\over {1\over y} + {y(1+i\epsilon)\over q^2} -
i\phi} - {1\over {1\over y} + {y(1+i\epsilon)\over q^2} +
i\phi} \right) \right]
\label{Z'2}
\ee
where $\phi \equiv VZ = b Z(q) q^{-4/3}$.

Since $\epsilon \sim 10^{-2}$ is a small number it may be neglected
and changing the integration variable, $y=qt$,
we come to the expression:
\be
Z'(q) = {10^{10}K(\sv)^2 \over 8\pi^2} q^2\chi (q) \sum_n (-1)^{n+1}
\int_0^\infty {dt\,t^2 \cos (nqt) \over (1+t^2)^2 + t^2\,\chi^2 (q)}
\label{Z'3}
\ee
where $\chi (q) = q\phi(q)= b Z(q) q^{-1/3}$. Both the integral over $t$
and summation over $n$ can be done explicitly and we finally obtain:
\be
\kappa\, Z' =  {10^{10}K(\sv)^2 \over 16\pi}
{ q^2 \over \sqrt{\chi^2 +4}} \left[ t_2 f_{eq} (qt_2) -
t_1 f_{eq} (qt_1) \right]
\label{Z'fin}
\ee
where we introduced the coefficient $\kappa$, such that $\kappa =1$
for large $q$ and $\kappa = 2$ for $q\sim 1$. It reflects the role
of decoherence terms, proportional to $\gamma$, see discussion after
eq.~(\ref{byparts}). The quantities $t_{1,2}$ are the poles of the
denominator in eq.~(\ref{Z'3}) in the complex upper half-plane of 
$t$ (resonances):
\be
t_{1,2} = {\sqrt{\chi^2 +4} \pm \chi \over 2}
\label{t12}
\ee
It is an ordinary differential equation that can be easily integrated
numerically. It quite accurately describes evolution of the lepton
asymmetry in the limit when back reaction may be neglected: we assumed
above that $\Sigma = 2$ and $\Delta \ll \Sigma$. 

Before doing numerical integration let us consider two limiting cases
of $q$ close to initial value when asymmetry is very small and the
case of large $q$. When $q$ is not too large, $q\sim 1$, the r.h.s. 
of eq.~(\ref{Z'fin})
can be expanded in powers of $\chi$ and we obtain a very simple 
differential equation that can be integrated analytically:
\be
Z'= {10^{10}K(\sv)^2 \over 64 \pi} q^2 f_{eq}(q) \chi(q)
\left[ -1 + q \left( 1- f_{eq}(q) \right)\right]
\label{Z'small}
\ee
(we took here $\kappa =2$). One sees that for 
$q< q_{min}=1.278$ 
the asymmetry exponentially decreases and reaches the minimum value
\be
{Z_{min} \over Z_{in}} = 
\exp \left[ -{10^{10} K(\sv)^2 b \over 64\pi}
\int_0^{q_{min}} dq \,q^{5/3} f_{eq}(q) \left(1 - {q\over 1+\exp(-q)}
\right)\right].
\label{zmin}
\ee
The integral in the expression above is equal to 0.07539 and e.g.
for $(\nue-\nus)$-oscillations the initial asymmetry drops by 3 orders
of magnitude in the minimum. The drop would be significantly stronger
even with a mild increase of mixing angle or mass difference.
The temperature, when the minimum is 
reached (corresponding to $q_{min} = 1.278$) is
\be
T_{min}^e = 17.3 \left( \dm \right)^{1/6}\,{\rm MeV},\,\,\,
T_{min}^{\mu,\tau} = 23.25 \left( \dm \right)^{1/6}\,{\rm MeV}   
\label{tmin}
\ee
These results rather well agree with ref.~\cite{dibari00c} for 
$\nue$, while agreement for $\num$ and $\nut$ 
case (see e.g. the papers~\cite{foot97b,volkas00,dibari00b})
is not so good.

For $q>q_{min}$ the asymmetry started to rise exponentially and this
regime lasted till $\chi$ becomes larger than one and the asymmetry 
reaches the magnitude $Z\sim 10^3$ or $L \sim 10^{-6}$. After
that the asymmetry started to rise as a power of $q$. 
For large $q$ and $\chi$ the term containing $t_2$ dominates the r.h.s.
of eq~(\ref{Z'fin}) and now it takes the form:
\be
Z^2\,Z'= {10^{10} K(\sv)^2 \over 16\pi \, b^2} q^{8/3} f_{eq} (1/VZ) 
\label{Zlarge}
\ee
where $V$ is given by eq.~(\ref{coef}). Assuming that $VZ$ is a slowly 
varying function of $q$ we can integrate this equation and obtain:
\be
Z(q) \approx 1.6\cdot 10^3 q^{11/9}\,\,\,
{\rm or}\,\,\, L(q) \approx 2.5\cdot 10^{-6}q^{11/9} 
\label{zlappr}
\ee
The concrete values of numerical coefficients above are taken for 
$(\nue-\nus)$-oscillations with $\dm =-1 {\rm eV}^2$ and $\sv= 10^{-4}$.
This result is in  a good agreement with the numerical solution of 
eq.~(\ref{Z'fin}) and the functional dependence, 
$L\sim q^{11/9}\sim T^{-11/3}$, agrees with that found in
ref.~\cite{buras00a} and slightly disagrees with the results of
refs.~\cite{foot96}-\cite{foot97b},\cite{dibari00b,dibari00c} 
where the law $L\sim q^{4/3}\sim T^{-4}$ was advocated. 

Numerical solution of eq.~(\ref{Z'fin}) is straightforward. It well 
agrees with the simple analysis presented above. In the power law 
regime, where the bulk of asymmetry is generated, it is accurately
approximated by the found above law~(\ref{zlappr}):
\be 
L_e = 2.5\cdot 10^{-6} C_e\, q^{11/9}
\label{Leofq}
\ee
For $\nue-\nus$ mixing with $(\sv)^2 = 10^{-8}$ and $\dm =-1$  
the correction coefficient $C_e$ is 0.96, 0.98, 1, 1.01, and 0.997
for $q=6630,\,\, 1000,\,\, 100,\,\, 10,$ and $5$ respectively.
The results of numerical solution well agree with those of 
ref.~\cite{dibari00c} in the temperature range from 10 down to 1 MeV
for $\nue-\nus$ case. 

For the $(\num-\nus)$-mixing with $\dm = -10$ and $(\sv)^2 = 10^{-9}$ 
the solution can be approximated as 
\be
L_\mu \approx 1.2\cdot 10^{-6} C_\mu \, q^{11/9}
\label{Lmuofq}
\ee
with the correction coefficient $C_\mu = 0.84,\,0.9,\, 0.98,\,1.02,\,
1.05,\, 1$   
for $q = 4\cdot 10^4,\, 10^4$, $10^3,\, 10^2,\,10,\,5 $ respectively.
These results reasonably well agree with the calculations of 
ref.~\cite{foot97b} in the
temperature range from 25 down to 2 MeV. At smaller temperatures
this power law generation of asymmetry must stop and it abruptly does, 
according to the results of the quoted papers, but the solutions 
of eq.~(\ref{Z'fin}) continue rising because back reaction effects 
are neglected there. We will consider these effects in the next 
section.

\section{Back reaction. \label{sec:br}}

The solution obtained above should be close to the exact one if
$\Delta (q,y) \ll \Sigma (q,y)$  and $\Sigma \approx 2 = const$,
see eq.~(\ref{Delta'}). Since now we know the function $Z(q)$ we 
can find $\Delta (q,y)$ and check when this assumptions are
correct. We will consider the region of sufficiently large $q$ 
when the second pole (resonance) $t_1$ in eq.~(\ref{Z'fin}) is
not important. Its contribution is suppressed as $\exp (-q^2VZ)$ 
and it may be neglected already at $q>5$ (we assume for 
definiteness that initial value of the asymmetry is positive, 
otherwise the role of the two poles would interchange).
In terms of oscillating coefficients $\cos \Delta\Phi$ or 
$\cos \bar\Delta\Phi$, entering eq.~(\ref{Delta'}), it means that 
only one of them is essential. It has a saddle point where the
oscillations are not too fast, while the other quickly oscillates
in all essential region of momenta $y$. With our choice of the
sign of the initial asymmetry only $\cos \bar\Delta\Phi$ has an 
essential saddle point and in this approximation the 
equation~(\ref{Delta'}) can be written as:
\be
\Delta' (q,y) = - { K^2 (\sv)^2 \over y^2 } \int_0^q dq_1 
\cos \bar\Delta\Phi
\label{Delta'2}
\ee  
For large $q$ the phase difference is equal to
\be
\Delta \Phi = K\left(-{q-q_1 \over y} + 
\int_{q_1}^q dq_2 V(q_2) Z(q_2) \right)
\label{DeltaPhi}
\ee
This integral can be taken in saddle point approximation. To this end
let us expand:
\be
\Phi (q,y) = \Phi (q_R,y) + {(q-q_R)^2\over 2} \Phi'' (q_R,y)
\label{Phiqy}
\ee
where the saddle (resonance) point $q_R$ is determined by the 
condition 
\be
\Phi' (q_R,y) = K\left( VZ - {1\over y}\right) = 0
\label{Phyres}
\ee
For $q<q_R$ the integral in the r.h.s. of eq.~(\ref{Delta'2}) may be
neglected, while for $q>q_R$ it is 
\be
\int_0^q dq_1 \cos \bar\Delta\Phi \approx 
\theta (q - q_R)\, {\cal R}e\,\left\{ \sqrt{{2\pi \over |\Phi''(q_R,y)|}}
\,\,e^{\left[ \Phi(q,y) -\Phi(q_R,y) -i\pi/4 \right] }\right\}
\label{intPhi}
\ee
where $\Phi'' = (VZ)'$.

Now repeating similar integration in eq.~(\ref{Delta'2}) we can easily 
find $\Delta (q,y)$:
\be
\Delta (q,y)= \theta(q-q_R){\pi\,K(\sv)^2\over y^2|(VZ)'|_R }
\label{Deltaqy}
\ee
Note the factor 1/2 that comes from the theta-function in 
expression~(\ref{intPhi}). It permits integration only over positive
values of $(q-q_R)$.

From the saddle point condition follows 
\be
(VZ)' = VZ \left( {V'\over V} + {Z'\over Z}\right)_R =
{1\over y}\left( {V'\over V} + {Z'\over Z}\right)_R =
-{1\over 9yq_R}
\label{VZ'}
\ee
In the last equality the solution $Z\approx 1.5\cdot 10^3\,q^{11/9}$ 
and $V=b q^{-4/3}$ were used. From the condition $(VZ)' = 1/y$ we find
\be
q_R \approx (5y)^9
\label{qr}
\ee
where we used  $\dm =-1$ and $b_e = 3.31\cdot 10^{-3}$.  

As a simple check we may calculate the integrated lepton asymmetry 
$L(q) = 16.45\cdot 10^{-10} Z(q)$:
\be
L(q)={16.45 \over 16\pi^2}\int_0^\infty dy y^2 f_{eq}(y)\Delta (q,y)
= {148K(\sv)^2 \over 16\pi}
\int_0^{y_{max}} dy\,y\, (5y)^{9} f_{eq}(y)
\label{Lq}
\ee
where $y_{max} = q^{1/9}/5$. This integral can be easily calculated 
and the result is in a good agreement with eq.~(\ref{zlappr}), as 
one should expect. 

However the magnitude of $\Delta (q,y)$ is too large. For example at
$q=6630$ (corresponding to $T = 1$ MeV for $(\nue-\nus)$-oscillations)
we find $\Delta = 9\pi K (\sv)^2 q_R /y \approx 200 \gg 1$. It 
violates the condition~(\ref{daslh}) and contradicts the assumption
that $\Delta \ll \Sigma$. Naively one would expect that the asymmetry
should be suppressed by two orders of magnitude but as we will see
below it is not the case. The evolution of $Z(q)$ changes, due to 
the back reaction, and the behavior of the resonance $y_R(q)$ also 
becomes different from $y_R = q^{1/9}/5$ found above. 

If only one resonance is essential then $(\Sigma +\Delta)$ is 
conserved, as is seen from eqs.~(\ref{Delta'},\ref{Sigma'}) 
with $\cos \Delta \Phi \rar 0$. 
It corresponds to conservation of the total leptonic charge
if oscillations are efficient only in neutrino (or antineutrino) 
channel. In this case we come to the equation:
\be
\Delta' (q,y) = - { K^2 (\sv)^2 \over y^2 } \int_0^q dq_1 
\cos \bar\Delta\Phi \left[ 1- \Delta (q_1,y)\right] 
\label{Delta'br}
\ee  
This integral can be taken in the same way as above in the saddle
point approximation and we obtain:
\be
\Delta (q,y)=\theta(q-q_R) \lambda \left[ 1- \Delta (q_R,y)\right]
\label{Deltaqy1}
\ee
where the last term describes the back-reaction and 
\be
\lambda = {\pi K (\sv)^2 \over y^2 |(VZ)\,'|_R}
\label{lambda}
\ee
The derivative of $VZ$ is taken over $q$ at $q = q_R (y)$ found from
the resonance condition $V(q_R)Z(q_R) = 1/y$. 

Since $\theta (0) =1/2$, we find
\be
\Delta (q,y) = {2\lambda \over 2+\lambda}\,\theta (q-q_R)
\label{Deltalam}
\ee
With this expression for $\Delta (q,y)$ we can find integrated asymmetry
\be
Z(q) ={10^{10}\over 16\pi^2} \int_0^{y_R} dy y^2 f_{eq}(y) \,
{2\lambda \over 2+\lambda}
\label{Zbr}
\ee
where $y_R(q) = 1/[V(q)Z(q)]$. This equation can be reduced to an
ordinary differential equation in the following way. Let us introduce 
the new variable
\be
\tau = {1 \over VZ}
\label{tau}
\ee
and consider the new unknown function $q= q(\tau)$. Correspondingly  
$ Z = q^{4/3}(\tau) /(b\tau)$. The derivative 
over $q$ should be rewritten as:
\be
{d(VZ) \over dq} = - {1\over \tau^2} \left({dq\over d\tau}\right)^{-1}
\label{dvzdq}
\ee 
Under the sign of the integral over $y$ one should take $\tau = y$,
while the upper integration limit is $y_{max} = \tau$. Now we can
take derivatives over $\tau$ of both sides of the equation~(\ref{Zbr})
and obtain:
\be
{d\over d\tau}\,\left[ {q^{4/3} (\tau) \over b\tau }\right]=
{10^{10}\over 16\pi^2}\, {2\lambda \over 2+\lambda}\,
\tau^2 f_{eq}(\tau).
\label{dqdtau}
\ee
where now $\lambda = \pi K (\sv)^2\, dq/d\tau$.
This is the final equation for determination of the integrated 
asymmetry with the account of the back reaction. As initial condition
we take the magnitude of asymmetry found from solution of 
eq.~(\ref{Z'fin}) at $q=5$. At this $q$ the back reaction is still 
small but already the regime of one pole dominance begins. Under the 
latter assumption the above equation~(\ref{dqdtau}) is derived. 

The numerical solution of eq.~(\ref{dqdtau}) is straightforward. 
In particular, if one neglects $\lambda$ 
with respect to 2 in the denominator of the r.h.s. of this equation, then
its numerical solution gives exactly the same result for the asymmetry
as has been found above from eq.~(\ref{Z'fin}). An account of back
reaction is not essential at high temperatures but it is quite important
in the temperature region of Big Bang Nucleosynthesis. In particular, at
$T=1$ MeV the asymmetry is $L= 0.0435$ and at $T=0.5$ MeV it is
$L = 0.25$. These values are approximately 3 times smaller than those 
found without back reaction. Asymptotic constant value of $L$ is reached 
at $T< 0.3$ MeV and is equal to 0.35. These numerical values are found
for electronic neutrinos with $\dm =-1$ and $(\sv)^2 = 10^{-8}$. The 
asymptotic value is approximately twice smaller than that presented
in the paper~\cite{dibari00c}, the same is true for the magnitude
of the asymmetry in the nucleosynthesis region as well. Another important
effect for BBN is the shape of the spectrum of electronic neutrinos
that may noticeably deviate from the simple equilibrium one given
by expression~(\ref{feq}) even with a non-zero chemical potential. It
will be considered elsewhere.  

For $\num$ or $\nut$ with $\dm = -10$ and $(\sv)^2 = 10^{-9}$ the 
asymmetry $L_\mu$ asymptotically tends to 0.237 in a good agreement 
with ref.~\cite{foot97b}. For non-asymptotic values of temperature
the corrected by back reaction asymmetry is 
$L_\mu=6.94\cdot 10^{-3}$ for $T=3$ MeV (1.24 smaller than 
non-corrected one), $L_\mu= 0.025$ for $T=2$ MeV (1.48 smaller),
and $L_\mu= 0.164$ for $T=1$ MeV (2.6 times smaller).  

There is an easy way to find the asymptotic constant value of the
asymmetry, $Z_0$ or $L_0$. To this end is convenient to use 
eq.~(\ref{Zbr}) with a constant $Z$:
\be
\lambda = {3\pi K (\sv)^2 b^{3/4} Z_0^{3/4} \over 4 y^{1/4}}=
70.85 (\dm)^{1/4} (\sv/10^{-4})^2 L_0^{3/4}
\label{lambdaz0}
\ee
This result is the same both for $\nue$ and $\nu_{\mu,\tau}$.

Since the upper limit of the integral over $y$ is 
$y_{max} = q^{4/3} /(bZ_0)\gg 1 $  for large $q$, we obtain the 
following equation for $L_0$:
\be
L_0 = 0.208 \int_0^\infty dy y^2 f_{eq}(y)\, 
{\lambda \over 2+\lambda}
\label{z0}
\ee
Numerical solution of this equation gives $L = 0.35$ for $\dm =-1$
and $(\sv)^2 = 10^{-8} $ and $L = 0.27$ for $\dm =-10$ and
$(\sv)^2 = 10^{-9} $ in a good agreement with the solution of
differential equation~(\ref{dqdtau}).

\section{Discussion.} \label{sec:discussion}

Thus, the statement of a generation of a large lepton
asymmetry by oscillations between active and sterile neutrinos  
is essentially confirmed here. The agreement between the present
calculations and those of ref.~\cite{foot97b} for the case of $\num$ and
$\nus$ mixing is very good, while there is a noticeable difference 
between the results of the present paper and the paper~\cite{dibari00c}
for the case of $\nue$ and $\nus$ mixing. In the temperature range
important for the primordial nucleosynthesis the results of the
present paper is 2-3 times smaller. 

There are some other minor differences. According to our results
the asymmetry rises as $T^{11/3}$, while the fit to numerical solution
of the papers quoted in the text is $T^4$. This difference is very 
important for the evolution of the resonance value of the neutrino
momentum $y$ and may possibly explain the difference of the results.
On the other hand, if this is the case, a good agreement for the 
$\num-\nus$ case becomes mysterious. 

Another difference between the present approach and the
calculations of 
refs.~\cite{foot96}-\cite{foot99},\cite{dibari00b,dibari00c}   
is the treatment of repopulation of active neutrino states. In those
papers the equilibrium number density of active neutrinos was taken
in the form~(\ref{feq}) with a non-zero chemical potential which value
is found in a self-consistent way together with calculation of the 
lepton asymmetry. This may be true when the decoherence effects,
described by $\gamma$, are strong. But the effects of neutrino production
were important only at high temperatures, when a minor fraction of 
asymmetry was generated. At that stage the contribution from non-zero
$\mu$ into kinetic equations are negligible in comparison with the terms
coming from $Z$ in effective potential, the latter are amplified by 
the factor $10^7$. Moreover, if the effects of $\mu$ were non-negligible,
one should take into account similar effects that lead to the difference
between $\gamma$ and $\bar\gamma$ of the same magnitude. The agreement
between the calculations for the $(\num-\nus)$ case discussed above shows
that one may neglect $\mu$ in the equilibrium distribution function.

The calculations presented here does not show any chaoticity. To be
more precise numerical solution of equation~(\ref{Z'fin}) shows 
chaotic behavior with an increasing $K\sv$. However this chaoticity
is related to numerical instability because with 
the increasing coefficient in front of the r.h.s. of the equation 
the minimal value of the asymmetry
becomes very small and can be smaller than the accuracy of computation.
In this case the numerically calculated value of the asymmetry may 
chaotically change sign. However this 
regime is well described analytically and it can be seen that the sign
of asymmetry does not change. There could be another effect first 
discussed in ref.~\cite{dibari00d} (see also~\cite{enqvist00}) 
- small primordial fluctuations of the cosmological (baryonic) charge 
asymmetry could be amplified by oscillations. This effect is related to 
neutrino diffusion and has nothing to do with the discussed chaoticity.
A chaotic behavior was observed in several recent 
papers~\cite{enqvist99,sorri00,braad00} in a simplified approach
when the kinetic equations were solved for a fixed ``average''
value of neutrino momentum, $y= 3.15$, so that integro-differential
equations become much simpler ordinary differential ones. However 
many essential features of the process could be obscured in this 
approach, in particular ``running of the resonance'' over neutrino 
spectrum and it is difficult to judge how reliable are the results. 
Moreover, the average value of $1/y$ that enters the refraction
index, is $\langle 1/y \rangle \approx 1$ and not 
$1/\langle y\rangle \approx 0.3$, though this numerical difference
might not be important for the conclusion.

Possibly fixed-momentum approach is not adequate
to the problem as can be seen from a very simple example. Let us 
consider oscillations between $\nua$ and $\nus$ in vacuum. Then
for a neutrino with a fixed energy the 
leptonic charge in active neutrino sector would oscillate with a 
very large frequency, $\sim \sin (\dm t /E)$. However if one averages
this result with thermal neutrino spectrum, the oscillations 
of leptonic charge would be exponentially suppressed. Still this 
counter-example is also oversimplified and cannot be considered
as a rigorous counter-argument. 
It may happen for example that in the case of smaller $K$
when saddle point does not give a good approximation, the differential
asymmetry $\Delta (q,y)$ may be an oscillating function but the
integrated (total) asymmetry is smooth. Another logically open option
is that for a smaller $K$ the asymmetry might be chaotic but not large.
Momentum dependent equations were analyzed in ref.~\cite{dibari00e}
and a chaoticity for some values of parameters were observed but
the authors could not exclude its origin by an instability of numerical 
computation procedure.
At this stage is difficult to make a final judgment.

As we have already mentioned the semi-analytic solution found in 
this paper confirms a strong generation of lepton asymmetry. The
results obtained are accurate in the limit of large values of 
parameter $K$ and for sufficiently small mixing, 
$(\sv)^2 < 0.01-0.001$.
For a larger mixing the population of sterile states becomes very 
strong and the $\nus$ and $\bar \nus$ states closely 
approach equilibrium
and become equally populated, so the asymmetry is not generated. For 
small values of the product $K\sv$ the process of asymmetry generation
is not efficient and the net result is rather low. Numerical 
calculations of the effect for very low values of the mass difference
$\dm = 10^{-7} - 10^{-11}$ eV${^2}$ show that the asymmetry could rise
only up to 4 orders of magnitude~\cite{kirilova97}-\cite{kirilova00}
producing the net result at the level of $10^{-5}$. According to
the calculations of this work, the asymmetry strongly rises if
$\dm > 10^{-3}$ eV$^2$ and possibly for smaller values depending upon
the mixing angle. More accurate estimates of the parameter range where 
the asymmetry may strongly rise and the role of the asymmetry generation 
in big bang nucleosynthesis will be discussed elsewhere.

\bigskip
{\bf Acknowledgement.} I am grateful to M. Chizhov, S. Hansen, D.
Kirilova,
and F. Villante for discussions and helpful comments.

\end{document}